\documentclass{article}

 \usepackage[final]{neurips_2025_ml4ps}

\usepackage[utf8]{inputenc} 
\usepackage[T1]{fontenc}    
\usepackage{hyperref}       
\usepackage{url}            
\usepackage{booktabs}       
\usepackage{amsfonts}       
\usepackage{nicefrac}       
\usepackage{microtype}      
\usepackage{xcolor}         

\usepackage[T1]{fontenc}
\usepackage[utf8]{inputenc}
\usepackage[utf8]{inputenc}
\usepackage[T1]{fontenc}
\usepackage{amsmath}
\usepackage{amsfonts}
\usepackage{amssymb}
\usepackage{geometry}
\usepackage{enumitem}
\usepackage{booktabs}
\usepackage{graphicx}
\usepackage{natbib}
\usepackage{tikz}
\usetikzlibrary{shapes.geometric, arrows.meta, positioning}
\usepackage{noto}
\definecolor{lightgray}{gray}{0.9} 

\title{FOXES: A Framework For Operational X-ray Emission Synthesis}

\author{
  Griffin T. Goodwin\thanks{Correspondence to: ggoodwin5@gsu.edu} \\
  Department of Physics \& Astronomy\\ 
  Georgia State University\\
  Atlanta, GA, USA\\
  \And
  Jayant Biradar\\
  College of Information Science\\
  University of Arizona\\
  Tucson, AZ, USA \\
  \And
  Alison J. March\\
  College of Engineering \& Applied Science\\
  University of Colorado\\
  Boulder, CO, USA \\
  \And
  Christoph Schirninger \\
  Institute of Physics\\
  University of Graz\\
  Graz, Austria \\
  \And
  Robert Jarolim \\
  High Altitude Observatory\\
  NSF National Center for Atmospheric Research\\
  Boulder, CO, USA \\
  \And
  Angelos Vourlidas \\
  The Johns Hopkins University Applied Physics Laboratory\\
  Laurel, MD, USA \\
    \AND
  Lorien Pratt \\
  Quantellia LLC\\
  Denver, CO, USA\\
}

\begin{document}

\maketitle

\begin{abstract}
  Understanding solar flares is critical for predicting space weather, as their activity shapes how the Sun influences Earth and its environment. The development of reliable forecasting methodologies of these events depends on robust flare catalogs, but current methods are limited to flare classification using integrated soft X-ray emission that are available only from Earth's perspective. This reduces accuracy in pinpointing the location and strength of farside flares and their connection to geoeffective events.
  In this work, we introduce a Vision Transformer (ViT)-based approach that translates Extreme Ultraviolet (EUV) observations into soft x-ray flux while also setting the groundwork for estimating flare locations in the future. The model achieves accurate flux predictions across flare classes using quantitative metrics. This paves the way for EUV-based flare detection to be extended beyond Earth’s line of sight, which allows for a more comprehensive and complete solar flare catalog.
\end{abstract}

\section{Introduction}

Solar flares are radiative manifestations of rapid magnetic energy release within the Sun's corona. Although solar flares rarely have a direct impact on Earth,
they serve as the best proxies for larger events such as coronal mass ejections (CMEs) and solar energetic particle events. When these events reach Earth, they have the potential to affect much of our core infrastructure, including satellites, radio communications, GPS, and power grids \citep{Schrijver2009, natras2019strong}. Consequently, it has become extremely important to develop accurate and reliable flare forecasting models to mitigate these risks \citep{barnes2016flare_workshop, Leka2019forecasting_methods}.  


One of the key ingredients necessary to enable these efforts is access to comprehensive and accurately labeled flare catalogs. However,  current catalogs face limitations, including gaps in data, errors in flare locations, and even uncertainties in flare strengths when multiple events occur simultaneously. Consequently, there remains significant room for improvement in this area, something that we aim to address within this work. 

Currently, the primary approach for the classification of solar flares is through the integrated SXR irradiance from the Sun, measured by the Geostationary Operational Environmental Satellite (GOES) \citep{woods2024}. Flares are categorized on a logarithmic scale provided by the National Oceanic and Atmospheric Administration (NOAA). The classes from weakest to strongest are as follows: \emph{A} (SXR flux $< 10^{-7}\,W/m^2$), \emph{B}, \emph{C}, \emph{M}, and \emph{X} ($\geq 10^{-4}\,W/m^2$).  Since the measured SXR value by GOES is the sum of the emission across the entire solar disk, this poses a significant limitation in terms of spatial localization for flares. \emph{Essentially, we cannot accurately determine individual flare intensities and locations from GOES alone.} Additionally, there are no SXR monitors (and none planned at this point) at locations away from the Sun-Earth line, \emph{severely curtailing our ability to protect human exploration of Mars from solar activity using existing methodologies}. 

To address these shortcomings, we present \emph{FOXES: A Framework for Operational X-ray Emission Synthesis}. Utilizing Vision Transformers (ViTs; \citealp{dosovitskiy2020image}), FOXES accurately learns the mapping between EUV observations and the integrated SXR flux measurements of GOES, allowing us to leverage the spatial information provided by the EUV data, to pin-point flare locations, while simultaneously extracting flare strength. In theory, this model can then be applied to any EUV-based spacecraft to estimate SXR flux, including on the far-side of the Sun.

In its current state, FOXES is only able to predict the \emph{integrated} SXR flux received at \emph{Earth}. However, with further development, FOXES will enable precise flare localization via patch-based SXR forecasting, allowing us to accurately quantify individual flare strengths for sympathetic (simultaneously occurring) flares. Additionally, through proper calibration, FOXES will be compatible with satellites such as the Solar TErrestrial RElations Observatory (STEREO).

In the long run, this work will contribute significantly to the forecasting of solar flares by providing additional flaring events and more accurate labels for benchmark training and testing datasets, especially during times of high flaring activity.

\section{Data}
In this work, EUV observations from the Solar Dynamics Observatory (SDO; \citealp{Pesnell2012}), taken by the Atmospheric Imaging Assembly (AIA; \citealp{Lemen2011}) are used as input to our FOXES model to determine the integrated SXR measurements provided by the GOES X-ray Sensor (XRS; \citealp{chamberlin2009}). For this translation, we focus on six EUV channels that are sensitive to flaring dynamics with 94~\AA, 131~\AA, 171~\AA, 193~\AA, 211~\AA, and 304~\AA, sampled at a one-minute cadence \cite{petkaki2012sdo}. To obtain a machine learning-ready dataset, we preprocess the EUV data using the Instrument-to-Instrument (ITI; \citealp{jarolim2025}) tool. This pipeline includes a crop of the full Sun observations to 1.1 solar radii, a correction for device degradation, an image and exposure normalization, as well as downsampling of images to a resolution of 512$\times$512 pixels. We divide our dataset into training, validation, and test sets, ensuring no temporal overlap between the data. An overview of the splits, along with the number of observations in each flare class, is provided in Table\,\ref{table:data}.

\begin{table}[h!tbp]
\caption{Summary of the number of observations in the training, validation, and test sets for the different flare classes. The dataset was temporally split as follows: \emph{Train} — Jan 2013--Dec 2022; Jul 1--20 2023, \emph{Validation} — Jan--Jun 2023; Jul 25--30 2023, \emph{Test} — Aug 2023--Sep 2025.} 
\label{table:data}      
\centering                          
\renewcommand{\arraystretch}{1.1} 
\begin{tabular}{|c||c c c|}        
\hline              
Flare Class & Train & Validation & Test \\  
\hline
\quad $<$C & 13,256 & 358 & 13,370 \\
\quad C & 48,140 & 15,722 & 63,503 \\
\quad M & 16,604 & 2,827 & 22,256 \\
\quad X & 1,738 & 81 & 1,075 \\
\hline                             
\end{tabular}
\end{table}
As ground truth, we rely on the 1-minute averaged GOES SXR measurements in the 1–8~\AA~band,  used for flare classification. In total, our dataset spans January 2012 to September 2025, with high cadence (1-minute) data captured between July 2023 to August 2023 and randomly sampled flaring and non-flaring data captured elsewhere.

\section{Methodology}

Vision Transformers are shown to be highly applicable for tasks such as image classification, object detection, semantic segmentation, and image generation \citep{vaswani2017}. Unlike traditional convolutional neural networks (CNNs), which focus on small fixed image patches, ViTs are able to capture relationships across the entire image. This capability is particularly valuable in our case, as it enables us to determine the spatial distribution of the integrated SXR emission. In addition, ViTs provide interpretable predictions thanks to their self-attention maps. This allows us to confirm that our model is focusing on physically relevant features within the EUV observations rather than random associations between the data.

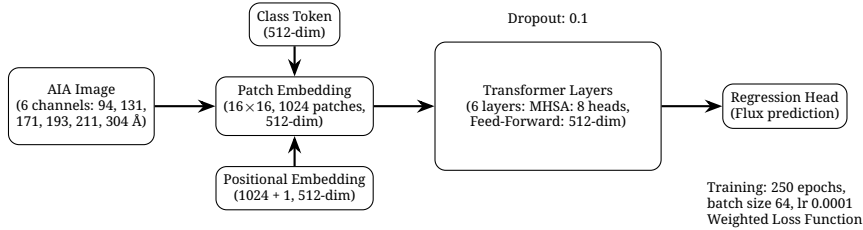
\begin{figure}[ht]
\centering 
\begin{tikzpicture}[
    box/.style={rectangle, draw, rounded corners, minimum height=0.6cm, minimum width=1.2cm, align=center, font=\tiny},
    arrow/.style={-Stealth, thick},
    font=\tiny,
    scale=0.85,
    every node/.style={scale=0.85}
]

\node[box, minimum width=2cm, minimum height=1.2cm] (input) at (0,0) {AIA Image\\(6 channels: 94, 131,\\171, 193, 211, 304 \AA)};

\node[box, right=0.8cm of input] (patches) {Patch Embedding\\(16$\times$16, 1024 patches,\\512-dim)};

\node[box, above=0.4cm of patches] (cls) {Class Token\\(512-dim)};
\draw[arrow] (cls) -- (patches.north);

\node[box, below=0.4cm of patches] (pos) {Positional Embedding\\(1024 + 1, 512-dim)};
\draw[arrow] (pos) -- (patches.south);

\node[box, right=0.8cm of patches, minimum height=2cm, minimum width=3.5cm] (transformer) {Transformer Layers\\(6 layers: MHSA: 8 heads,\\ Feed-Forward: 512-dim)};



\node[box, right=0.8cm of transformer] (output) {Regression Head\\(Flux prediction)};
\draw[arrow] (input) -- (patches);
\draw[arrow] (patches) -- (transformer);
\draw[arrow] (transformer) -- (output);

\node[above=0.1cm of transformer] {Dropout: 0.1};
\node[below=0.6cm of output, align=left] {Training: 250 epochs,\\batch size 64, lr 0.0001\\Weighted Loss Function};

\end{tikzpicture}

\caption{Our FOXES architecture, highlighting patch embedding, class token, positional embedding, transformer layers with multi-head self-attention (MHSA), and output head.}
\label{fig:vit_architecture}
\end{figure}

Utilizing a modified version of PyTorch Lightning's \citep{lightningai_vit_2025} ViT implementation (see Figure \ref{fig:vit_architecture}), our model processes the six EUV wavelength channels (94, 131, 171, 193, 211, 304 \AA) simultaneously to predict a single SXR output for a given timestamp. Through experimentation, we chose a 16$\times$16 pixel patch size, to create 1024 patches to balance detail and computational efficiency. Patches are embedded into a 512-dimensional space to capture multi-wavelength information. The model uses six layers with eight attention heads and a 512-unit feed-forward layer for efficient processing. A dropout layer of 0.1 was included to prevent overfitting, and a learning rate of 1e-4 with cosine annealing to a learning rate of 1e-5 ensures stable training over 250 epochs with a batch size of 64. Huber loss was utilized during training and was weighted by the inverse frequency of each flare class (flare quiet, C, M, X) in the training dataset. This ensured that class imbalance was mitigated during training. In general, these design choices prioritize the capture of complex solar features, while maintaining practical performance.

\section{Results}
\begin{figure}
  \centering
    \includegraphics[width=.8\textwidth]{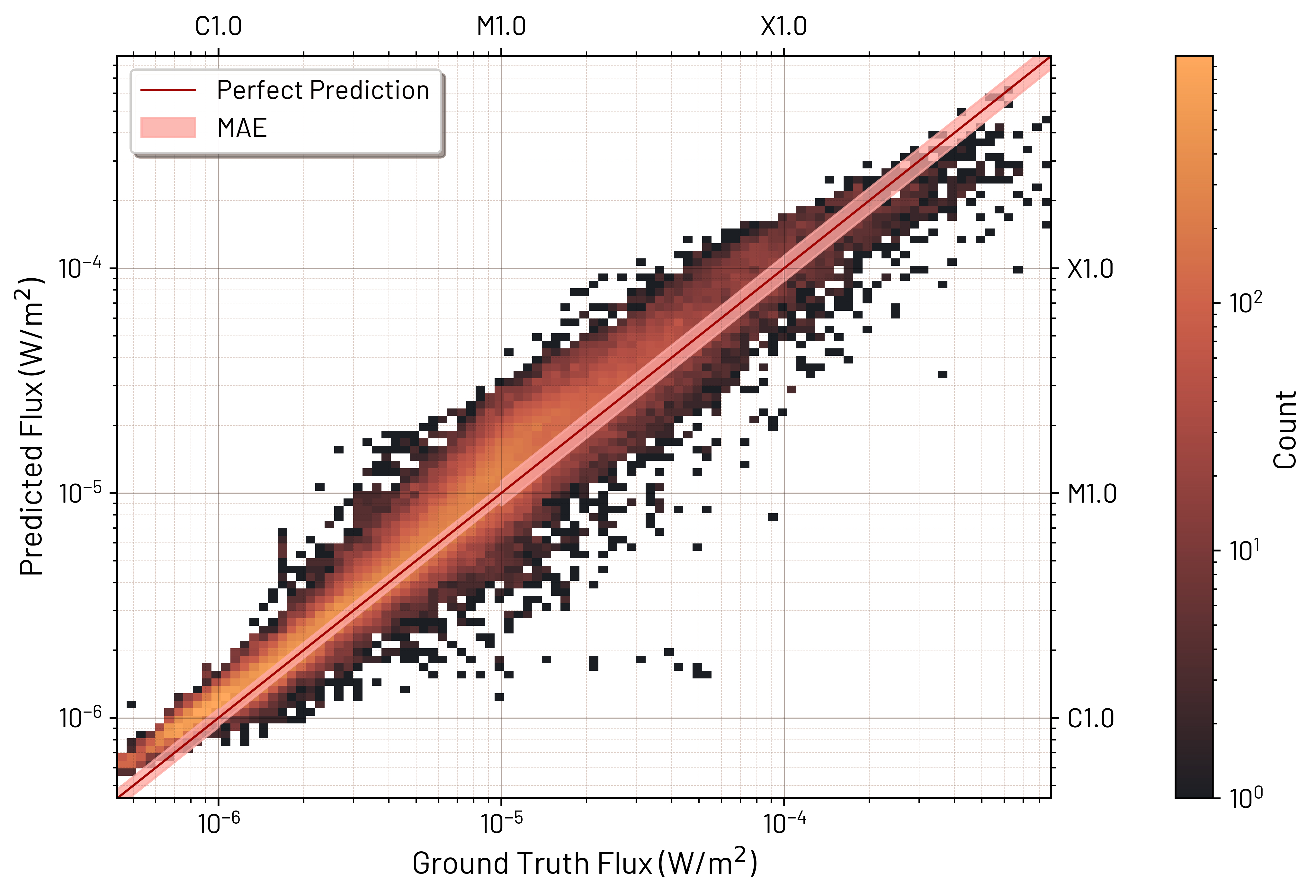}
  \caption{A 2D histogram comparing our FOXES model performance to the ground truth. A perfect prediction line (dark red) and the mean absolute error (in log-space) of our model at different flare classes (shaded red) is overlaid on top.}
  \label{regression}
\end{figure}

Figure \ref{regression} shows the results of our FOXES model applied to the test dataset. The figure presents a 2D histogram comparing the predicted SXR flux from our FOXES model versus the ground truth provided by GOES. We overlay a perfect prediction line (dark red), along with the mean absolute error (MAE; calculated in log-space) of our model at different flare classes. It is clear from the graph that most predictions lie relatively close to the one-to-one line, with FOXES performing particularly well for SXR measurements at the C-class level, as can be seen from the low MAE. In addition, both M- and X-class data points show relatively good agreement with the ground truth, as our predictions typically fall within a few magnitudes in flux. Nevertheless, there is room for improvement. Future work will focus on additional hyperparameter tuning to further enhance our results. Additional quantitative metrics are summarized in Table \ref{table:qualitymetrics}.

\begin{table}[h!tbp]
\caption{A summary of the quantitative metrics calculated in log-space for our FOXES model. Mean squared error (MSE), root mean squared error (RMSE), mean absolute error (MAE), and Pearson correlation coefficient (r) are shown for the overall performance of the model, as well as individual flare classes (this includes data both during background emission and flaring events, as long as they meet the appropriate flux thresholds).} 
\label{table:qualitymetrics}      
\centering 
\renewcommand{\arraystretch}{1.1} 
\begin{tabular}{|c||c c c c|}        
\hline              
Flare Class & MSE &  RMSE &  MAE  & r \\       
\hline
\textbf{Overall} & \textbf{1.41e-2} & \textbf{1.19e-1} & \textbf{8.97e-2} & \textbf{0.982} \\
\hline
\quad $<$C & 9.52e-3 & 9.76e-2 & 9.22e-2 &  0.905 \\
\quad C & 1.07e-2 & 1.03e-1 & 7.56e-2 &  0.962 \\
\quad M & 2.56e-2 & 1.60e-1 & 1.26e-1 &  0.857 \\
\quad X & 3.19e-2 & 1.79e-1 & 1.29e-1 &  0.665 \\
\hline                             

\end{tabular}
\end{table}

\begin{figure}
  \centering
    \includegraphics[width=1\textwidth]{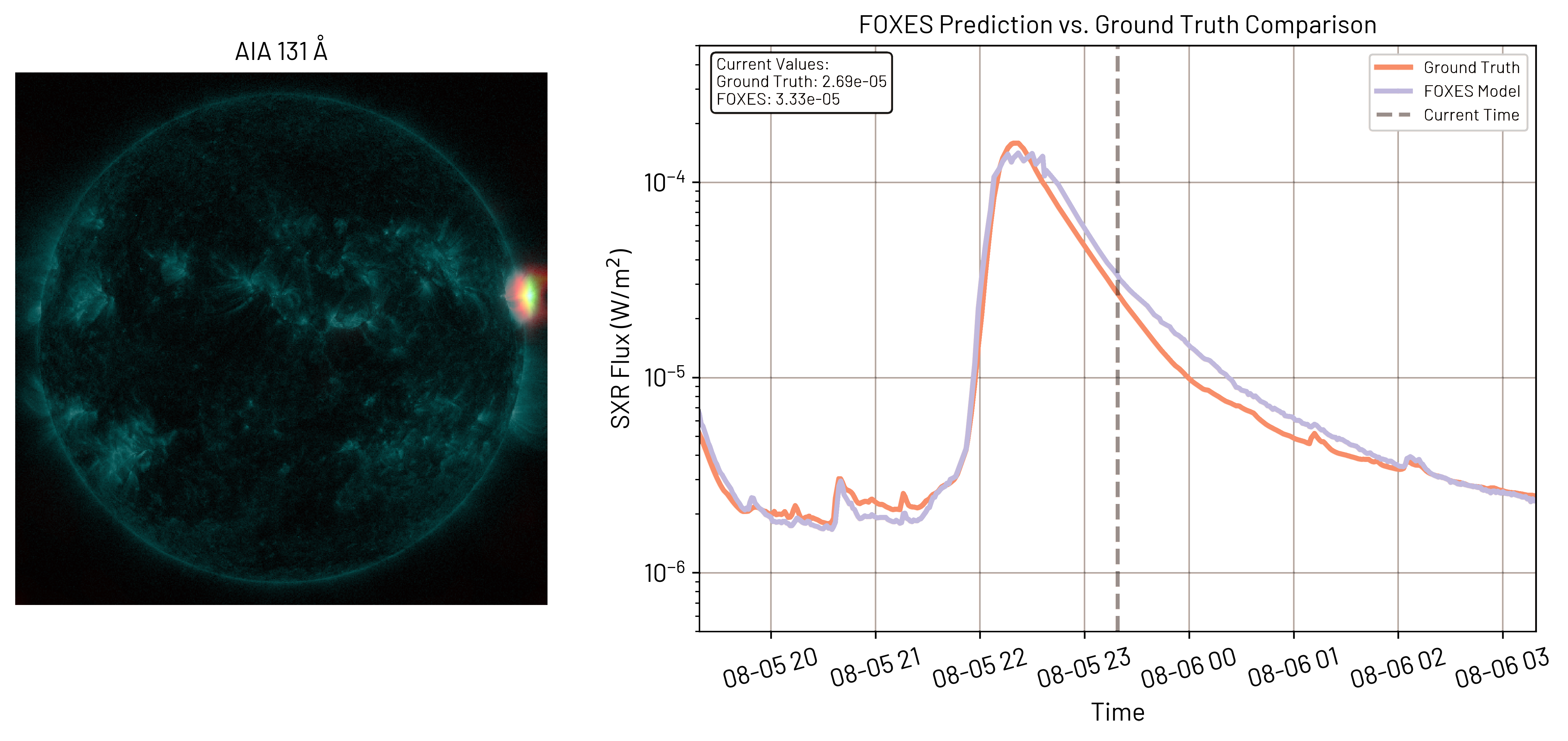}
  \caption{(Left Panel) An SDO/AIA 131~\AA~image, with attention-based heat map, provided by FOXES, overlaid in red. (Right Panel) A comparison of our FOXES model to the ground truth for an X-class flare that occurred on August 5th, 2023. The timestamp of the left image is shown by the gray dotted line.}
  \label{attention}
\end{figure}

Figure \ref{attention} illustrates the capability of our FOXES model to estimate SXR measurements. The image in the left panel shows the attention map provided by our model, overlaid on the 131 ~\AA~ AIA channel (a particularly sensitive channel to flaring events). In the right panel, we can see the SXR translation provided by FOXES for a flaring event in August of 2023, along with the time corresponding to the image. What is most impressive is that FOXES captures not only the impulsive and peak phases of the event, but also its decay phase. In addition to this, FOXES provides meaningful attention maps, aligned with observed active regions, strongly highlighting the actual flaring region on the top right of the image. This suggests that FOXES captures some of the physical relationships necessary to link EUV observations to SXR measurements.




\section{Conclusion}
In summary, FOXES demonstrates that multichannel EUV images can reliably estimate the integrated GOES SXR flux. The success of this proof of concept opens several exciting avenues to explore in the future. For example, missions traveling outside the Sun-Earth line, such as the Solar TErrestrial RElations Observatory (STEREO) and Solar Orbiter, provide multichannel EUV images of the Sun, but cannot directly measure the integrated SXR flux. Consequently, the identification and classification of flares on the far-side of the Sun is challenging, at best. As a result, astronauts and satellite missions that may be traveling outside the Sun-Earth line in the future are at great risk. By applying our FOXES model to ITI calibrated data \citep{jarolim2025} from these missions, we can provide more reliable estimates of SXR emission. As an added benefit, the new flare strengths that we collect will help to create a more comprehensive and reliable flare catalog, which will further improve current flare forecasting algorithms and provide better statistics on flare occurrence. Lastly, to further enhance our work, FOXES could be extended into a patch-based model, predicting SXR flux for individual sub-regions to enable more accurate localization and characterization of simultaneous flares. If successful, this advancement would represent a significant step forward for the space weather community.


\section*{Acknowledgments}
This work is a research product of Heliolab (heliolab.ai), an initiative of the Frontier Development Lab (FDL.ai). FDL is a public–private partnership between NASA, Trillium Technologies (trillium.tech), and commercial AI partners including Google Cloud and NVIDIA.
Heliolab was designed, delivered, and managed by Trillium Technologies Inc., a research and development company focused on intelligent systems and collaborative communities for Heliophysics, planetary stewardship and space exploration.
We gratefully acknowledge Google Cloud for extensive computational resources, and NVIDIA Corporation.
This material is based upon work supported by NASA under award number No. 80GSFC23CA040. Any opinions, findings, and conclusions or recommendations expressed are those of the author(s) and do not necessarily reflect the views of the National Aeronautics and Space Administration.


\bibliographystyle{plainnat}
\bibliography{references.bib}

\end{document}